\documentclass[12pt,a4paper]{iopart}

\usepackage{iopams}
\usepackage{setstack}
\usepackage{graphicx}

\usepackage{epstopdf}

\usepackage{xspace}

\newcommand{\sra}[1][]{\ensuremath{\quad \mathop{\longrightarrow}_{#1} \quad}}

\newcommand{\evat}[3][]{\ensuremath{\left.#2\right|^{#1}_{#3}}}
\newcommand{\deriv}[3][]{\ensuremath{\frac{\rmd^{#1} #2}{{\rmd #3}^{#1}}}}
\newcommand{\pderiv}[3][]{\ensuremath{\frac{\partial^{#1} #2}{{\partial #3}^{#1}}}}
\newcommand{\hderiv}[3][]{\ensuremath{\rmd^{#1} #2/{\rmd #3}^{#1}}}

\newcommand{\derivat}[4][]{\ensuremath{\evat{\deriv[#1]{#2}{#3}}{#4}}}

\newcommand{\derivateq}[4][]{\ensuremath{\derivat[#1]{#2}{#3}{#3=#4}}}

\newcommand{\hfrac}[2]{\ensuremath{\left.#1\middle/#2\right.}}

\newcommand{\fr}[1]{\ensuremath{\frac{1}{#1}}}

\newcommand{\ahalf}{\ensuremath{\fr{2}}}
\newcommand{\athird}{\ensuremath{\fr{3}}}

\newcommand{\cte}{\ensuremath{\rm{const}}}

\newcommand{\ub}{\ensuremath{\bar{u}}}
\newcommand{\ks}{\ensuremath{\kappa_{\rm{st}}}}
\newcommand{\kp}{\ensuremath{\kappa_{\rm{st-per}}}}
\newcommand{\Uh}{\ensuremath{U_{\rm{H}}}}
\newcommand{\vl}{\ensuremath{v_l}}
\newcommand{\apr}{\ensuremath{a_{\rm{p}}}}

\newcommand{\rl}{\ensuremath{r_l}}
\newcommand{\tl}{\ensuremath{t_l}}
\newcommand{\xl}{\ensuremath{x_l}}
\newcommand{\rc}{\ensuremath{r_{\rm{c}}}}
\newcommand{\observer}{\ensuremath{\mathcal{O}}\xspace}

\begin{document}

\title[Hawking radiation: An analytic expression for the effective-temperature function]{Hawking radiation as perceived by different observers: An analytic expression for the effective-temperature function}

\author{L~C~Barbado$^1$, C~Barcel\'o$^1$ and L~J~Garay$^{2, 3}$}

\address{$^1$ Instituto de Astrof\'{\i}sica de Andaluc\'{\i}a (CSIC), Glorieta de la Astronom\'{\i}a, 18008 Granada, Spain}
\address{$^2$ Departamento de F\'{\i}sica Te\'orica II, Universidad Complutense de Madrid, 28040 Madrid, Spain}
\address{$^3$ Instituto de Estructura de la Materia (CSIC), Serrano 121, 28006 Madrid, Spain}

\eads{\mailto{luiscb@iaa.es}, \mailto{carlos@iaa.es}, \mailto{luisj.garay@fis.ucm.es}}

\begin{abstract}

Given a field vacuum state in a black hole spacetime, this state can be analysed in terms of how it is perceived (in terms of  particle content) by different observers. This can be done by means of the effective-temperature function introduced by Barcel\'o {\em et al.\/} in~\cite{Barcelo:2010pj}. In Barbado {\em et al.\/}~\cite{Barbado:2011dx}, this function was analysed in a case by case basis for a number of interesting situations. In this work, we find a general analytic expression for the effective-temperature function which, apart from the vacuum state choice, depends on the position, the local velocity and the acceleration of the specific observer.
We give a clear physical interpretation of the quantities appearing in the expression, and illustrate its potentiality with a few examples. 

\end{abstract}


\submitto{\CQG}


\section{Introduction}

In a recent paper~\cite{Barbado:2011dx}, the present authors studied how Hawking radiation~\cite{Hawking:1974sw} emitted by a black hole would be perceived by different observers in the spacetime. The vacuum state for the quantum radiation field in a stationary spacetime, a simple massless scalar field, was chosen non-stationary so that it  mimicked  the switching-on of Hawking radiation, as would happen in a realistic collapse scenario. Then, the perception of this radiation state was analysed by using the method introduced in~\cite{Barcelo:2010pj}, and further described in~\cite{Barcelo:2010xk}. This method is based in the introduction of an \emph{effective-temperature function} which depends on the observer state of motion\footnote{This effective-temperature function is based on null-ray behaviour, and so it is only relevant for massless fields.}. Under certain conditions, the radiation perception can be shown to have a thermal shape, with a temperature proportional to this effective-temperature function. Other approaches to the problem of radiation perception could be the direct calculation of the Bogoliubov coefficients~\cite{birrell1984quantum}, the functional Schr\"odinger formalism~\cite{Greenwood:2008zg}, or the response function of Unruh-DeWitt detectors~\cite{Louko:2007mu}. 

In this work, we construct an analytic expression for the effective-temperature function valid for any observer (under the restrictions of spherical symmetry). The expression depends firstly on the choice of vacuum state via what we shall call the \emph{state effective temperature.} For instance, for a Schwarzschild black hole in the Unruh vacuum state this state effective temperature turns out to be nothing but Hawking surface gravity $\kappa_{\rm{H}}=1/(4m)$. Then, the observer characteristics enter into the expression by means of its instantaneous position, local velocity and proper acceleration. As we will discuss, in terms of this physical quantities the effective-temperature function has a very simple form and a compelling physical interpretation.  

In~\sref{temp_function}, we briefly recall the definition of the effective-temperature function and its physical meaning. Then, we split this function into a state effective-temperature function and observer dependent factors. We also give some examples of state effective-temperature functions for different vacuum choices. In~\sref{expression}, we write down the general analytic expression for the effective-temperature function and show how it is derived. We follow with a detailed discussion of its physical interpretation. Lastly, in~\sref{applications} we illustrate how the obtained formula applies to different cases.

\section{The effective-temperature function\label{temp_function}}

Throughout this work we particularize our discussion to a Schwarzschild black hole, although it would be easy to adapt our expression to more general spherically symmetric spacetimes. In Schwarzschild coordinates the Schwarzschild metric reads ($G=c=1$)
\begin{equation}
\rmd s^2 = -\left(1-\frac{2m}{r}\right) \rmd t^2 + \left(1-\frac{2m}{r}\right)^{-1} \rmd r^2 + r^2 \rmd \Omega^2.
\label{metric}
\end{equation}
In this spacetime we assume that we  have a quantum massless scalar field in a particular (maybe non-stationary) vacuum state $|0\rangle$ with radiation characteristics. Defining an outgoing null coordinate as
\begin{equation}
\ub := t - r^*, \quad r^* := r + 2m \log \left( \frac{r}{2m} - 1 \right),
\label{ub_def}
\end{equation}
the vacuum state choice can be encoded in the selection of a new null coordinate $U$ given by the function $U=U(\ub)$ (see~\cite{Barbado:2011dx} for details). Starting from this function one could calculate, for instance, the particle content of this state at infinity. 

Once a vacuum state has been selected, different observers will perceive it differently. To analyse this vacuum perception, one can take a specific observer timelike trajectory $\left(t(\tau),r(\tau)\right)$, with $\tau$ being the proper time, and introduce this proper time as another null variable after a suitable synchronization: $u=\tau-\tau_0$. Then, we can construct the relation $\ub=\ub(u)$  using~\eref{ub_def}, and composing it with $U=U(\ub)$, obtain the relation $U=U(u)$. From this relation one can compute the particle content perception  associated with  this generic but specific observer, as it is the only ingredient needed for calculating the Bogoliubov coefficients~\cite{Barbado:2011dx}. It is from this relation $U(u)$ that the effective-temperature function for this observer is defined:
\begin{equation}
\kp (u) := - \hfrac{\deriv[2]{U}{u}}{\deriv{U}{u}},
\label{kappa_def}
\end{equation}
where ``st-per'' stands for ``state perception''. 
When $u$ corresponds to the future null coordinate in a black hole geometry, this function encodes the ``peeling'' of null geodesics in the geometry, which is the relevant notion for calculating the Hawking temperature of slowly evaporating horizons~\cite{Barcelo:2010pj,Barcelo:2010xk}. In our case, the $\kp (u)$ provides a notion of peeling of geodesics which incorporates information, not only about the geometry, but also about the specific vacuum state and observer.

\subsection{Interpretation of $\kp$ as the temperature of a thermal radiation}
As described thoroughly in~\cite{Barcelo:2010pj, Barcelo:2010xk}, when the function $\kp(u)$ is nearly constant $\kp(u) \simeq \kp (u^*)=:\kappa_*$ over a sufficiently large interval around a given $u^*$, one can assure that during this same interval the  observer is detecting a Hawking flux of particles with a temperature $T$ given by
\begin{equation}
k_{\rm B} T = \frac{\kappa_*}{2\pi},
\label{hawking_temperature}
\end{equation}
(we always talk about $\kappa$ as a temperature, forgetting about the proportionality constant). If the variation of $\kp(u)$ is slow, then the observer will detect a thermal radiation with a slowly varying temperature. Whether $\kp(u)$ varies slowly or not in the surroundings of $u^*$, is controlled by an adiabatic condition which, under mild technical assumptions~\cite{Barcelo:2010xk}, reads
\begin{equation}
\epsilon_* := \fr{\kappa_*^2} \left| \derivateq{\kp}{u}{u^*} \right| \ll 1.
\label{adiabatic_condition_alt}
\end{equation}
Only when this condition is satisfied, we can say that $\kp$ is  providing  the temperature of the  thermal radiation  perceived by the observer using the formula~\eref{hawking_temperature}.  When this is not the case, the value of $\kp$ does not represent strictly a temperature, but  is still an estimator of the overall rate of particle detections.

\subsection{The state effective-temperature function
\label{kappa_state}}

Once the vacuum state has been chosen via the relation $U(\ub)$, the effective-temperature function in~\eref{kappa_def} can be rewritten (by using the chain rule) as
\begin{eqnarray}
\kp (u) & = \left(-\hfrac{\deriv[2]{U}{\ub}}{\deriv{U}{\ub}}\right) \deriv{\ub}{u} - \hfrac{\deriv[2]{\ub}{u}}{\deriv{\ub}{u}} \nonumber \\
& = \deriv{\ub}{u} \ks(\ub) - \hfrac{\deriv[2]{\ub}{u}}{\deriv{\ub}{u}},
\label{kappa_rew}
\end{eqnarray}
where
\begin{equation}
\ks(\ub) := -\hfrac{\deriv[2]{U}{\ub}}{\deriv{U}{\ub}}~
\label{ks_def}
\end{equation}
is what we call the \emph{state effective temperature.} It is the part of expression \eref{kappa_rew} that contains the information about the definition of the state. The rest does not depend on the relation $U(\ub)$, and thus will depend only on the observer's trajectory. In fact, \eref{ks_def} fits with the definition of $\kp (u)$ in~\eref{kappa_def} if we replace $\ub=u$, which is actually true for an observer at rest at infinity. Therefore, $\ks$ is the value of $\kp$ that this particular observer would perceive. It is worth mentioning that the fact that $\ks$ only depends on $\ub$ shows that its value propagates with the outgoing light rays (it is constant along each outgoing light ray), which is relevant for non-stationary vacuum states.

\begin{figure}[ht]
	\centering
	\includegraphics{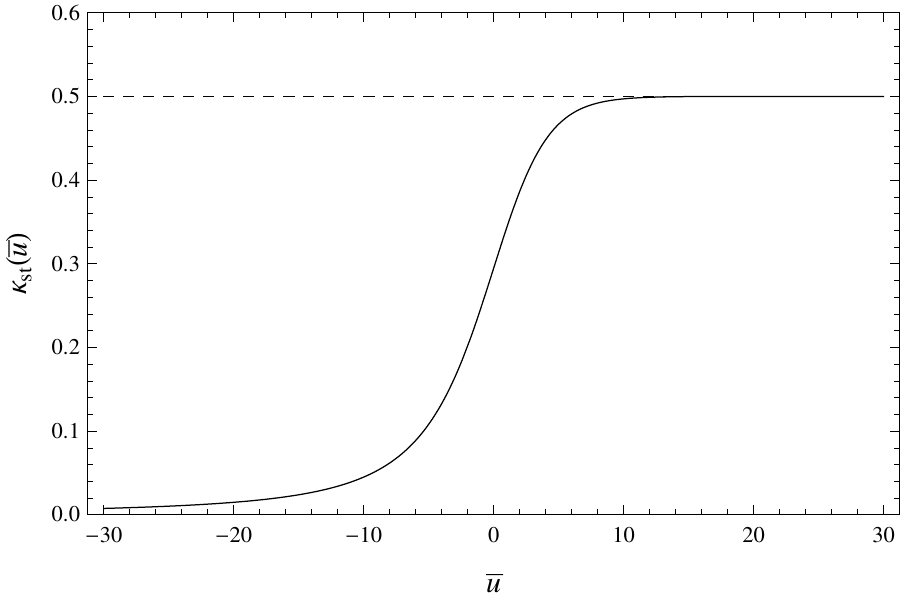}
	\caption{$\ks(\ub)$ in~\eref{non-stat}, for the non-stationary vacuum state (curve depicted \full), and $\ks = 1/(4m)$ for the Unruh vacuum state (curve depicted \dashed). We use $2m = 1$ units.\label{kappa_st}}
\end{figure}

We can analyse the form of this function for different vacuum-state choices. For the Boulware vacuum state~\cite{Boulware:1974dm}, we have $U(\ub) = \ub$ and thus $\ks(\ub) = 0$, which means that this vacuum does not radiate. For the Unruh vacuum state~\cite{Unruh:1976db,Fabbri:2005mw}, we have $U(\ub) = -4m \rme^{-\ub/(4m)}$ so that $\ks (\ub) = 1/(4m)$ (Hawking temperature). In the case of the non-stationary state defined in~\cite{Barbado:2011dx}, corresponding to that defined  in terms of positive frequency modes as seen by a free-falling observer from infinity, it is not difficult to obtain that
\begin{equation}
\ks(\ub) = \fr{4m \left[\frac{3}{4m}\left(\Uh - U(\ub)\right) + 1 \right]^{4/3}},
\label{non-stat}
\end{equation}
where $U(\ub)$ is known only through the following implicit relation:
\begin{eqnarray}
\fl \ub = \Uh+\frac{4m}{3}-4m\left\{\left[\frac{3}{4m}(\Uh-U)+1\right]^{1/3} + \ahalf \left[\frac{3}{4m}(\Uh-U)+1\right]^{2/3} \right. \nonumber \\
\left. +\athird \left[\frac{3}{4m}(\Uh-U)+1\right] + \log\left[\left(\frac{3}{4m}(\Uh-U)+1\right)^{1/3} -1\right]\right\},
\label{ub_of_U}
\end{eqnarray}
$\Uh$ being a constant labelling the event horizon. Noticing that $U(\ub)$ runs from $-\infty$ in the past infinity, to $\Uh$ in the future infinity, one can see that~\eref{non-stat} interpolates smoothly between $0$ and $1/(4m)$, as shown in figure~\ref{kappa_st}. Thus, this vacuum state qualitatively mimics the switching-on of Hawking radiation in realistic collapse scenarios.

A different non-stationary vacuum state is used in~\cite{Barbado:2011ai}. In this case, the state is designed to reproduce the radiation that would be emitted from a hypothetical stellar object supporting itself outside but extremely close to its gravitational radius, when experiencing small pulsations. The result found is that overall $\ks(\ub) \simeq 1/(4m)$. This means that the state is almost indistinguishable from the Unruh state when seen from infinity.

\section{General analytic expression for the effective-temperature function
\label{expression}}

Let us consider any observer~\observer outside the black hole following a trajectory $(t(u),r(u))$, $u$ being its proper time. The central result of this work is that the value of the effective-temperature function along the trajectory of \observer  can be expressed as
\begin{equation}
\kp (u) = \sqrt{\frac{1-\vl}{1+\vl}} \fr{\sqrt{1-\frac{2m}{r}}} \left(\ks(\ub)-\frac{m}{r^2}\right) + \apr.
\label{local_kappa}
\end{equation}
Here $r = r(u)$ and $\ub = \ub(u)$; the function $\vl = \vl(u)$ is the velocity of the observer~\observer with respect to the black hole as measured by a \emph{locally Minkowskian observer;} and the function $\apr = \apr(u)$ is the \emph{proper acceleration} of the observer~\observer, that is, the acceleration as measured by a \emph{free-falling observer instantaneously comoving with the observer~\observer.} Before discussing the clear physical interpretation of this expression, let us show in a constructive manner how it can be found.

\subsection{Calculation of the expression}

To show that expression~\eref{local_kappa} is indeed correct let us first describe the radial component of the trajectory $r(u)$ locally around the instant $u=u_0$. Up to second order in $u$, the trajectory of \observer can be described as
\begin{equation}
r(u) = r_0 + v_0 (u-u_0) + \ahalf a_0 (u-u_0)^2 + O(u-u_0)^3,
\label{local_traj}
\end{equation}
where
\begin{equation}
r_0 := r(u_0), \quad v_0 := \derivateq{r}{u}{u_0}, \quad a_0 := \derivateq[2]{r}{u}{u_0}.
\label{local_param}
\end{equation}

Using this local description of the trajectory and the line element~\eref{metric}, one can obtain $\hderiv{r}{u}$, $\hderiv[2]{r}{u}$, $\hderiv{t}{u}$, and $\hderiv[2]{t}{u}$ as local functions of $u$ around $u=u_0$. Then, using~\eref{ub_def} and~\eref{kappa_rew}, one can also obtain the local expression of $\kp (u)$ around $u=u_0$. The computation yields
\begin{eqnarray}
\fl \kp (u) =\fr{\sqrt{1-\frac{2m}{r_0}}} \left(\frac{\sqrt{1-\frac{2m}{r_0}+v_0^2}-v_0}{\sqrt{1-\frac{2m}{r_0}+v_0^2}+v_0}\right)^{1/2} \left(\ks(\ub(u_0))-\frac{m}{r_0^2}\right) \nonumber \\
+ \fr{\sqrt{1-\frac{2m}{r_0}+v_0^2}} \left(a_0 + \frac{m}{r_0^2} \right) + O(u-u_0).
\label{kappa_alt}
\end{eqnarray}
Now, it is not difficult to see that, if we identify
\begin{equation}
\vl := \frac{v_0}{\sqrt{1-\frac{2m}{r_0}+v_0^2}},
\label{vs_def}
\end{equation}
and
\begin{equation}
\apr := \fr{\sqrt{1-\frac{2m}{r_0}+v_0^2}} \left(a_0 + \frac{m}{r_0^2} \right),
\label{af_def}
\end{equation}
then~\eref{kappa_alt} is precisely~\eref{local_kappa} evaluated at $u=u_0$. However, as $u_0$ is \emph{any arbitrary instant} of the trajectory, we can conclude that the expression is valid all along the trajectory of the observer~\observer. Thus, it only remains to be shown that both definitions~\eref{vs_def} and~\eref{af_def} physically correspond to the described quantities at $u=u_0$.

Given a metric, there always exist coordinate systems such that in a particular, but arbitrary, point the metric becomes Minkowskian. Take the point $(t_0,r_0)$ in the metric~\eref{metric}. The transformation 
\begin{equation}
\left( \begin{array}{c}
	\rmd \tl \\ \rmd \rl
\end{array} \right)
=T \left( \begin{array}{c}
	\rmd t \\ \rmd r
\end{array} \right)
\label{static}
\end{equation}
between any locally Minkowskian coordinate system $(\tl,\rl)$ and the original one  $(t,r)$ has to satisfy
\begin{equation}
\evat{T}{u=u_0} =
\left( \begin{array}{cc}
	\gamma & -\gamma \vl \\ -\gamma \vl & \gamma
\end{array} \right)
\left( \begin{array}{cc}
	\sqrt{1-\frac{2m}{r_0}} & 0 \\ 0 & \frac{1}{\sqrt{1-\frac{2m}{r_0}}}
\end{array} \right),
\label{first_order}
\end{equation}
where  $\gamma:=1/\sqrt{1-\vl^2}$ is the usual relativistic boost factor. Here $\vl$ represents the Minkowskian velocity of the local coordinate system with respect to the static black hole (remember that in Schwarzschild coordinates the staticity of this spacetime is explicit). 

Now take the local Minkowskian coordinate system to be comoving with the trajectory $r(u)$
of the observer~\observer at $u_0$. In that case we have 
\begin{equation}
0=\derivat{\rl}{\tl}{u=u_0} = \frac{ v_0 -\vl \left(1 -\frac{2m}{r_0}\right)\derivat{t}{u}{u=u_0}}{\left(1 -\frac{2m}{r_0}\right)\derivat{t}{u}{u=u_0} -\vl v_0}.
\label{comoving}
\end{equation}
Directly from the metric~\eref{metric}, we known that
\begin{equation}
\derivat{t}{u}{u=u_0}=\frac{\sqrt{1 -\frac{2m}{r_0}+v_0^2}}{1 -\frac{2m}{r_0}}.
\end{equation}
Therefore, from~\eref{comoving} we obtain the expression 
\begin{equation}
\vl = \frac{v_0}{\sqrt{1-\frac{2m}{r_0}+v_0^2}}
\label{vel_static}
\end{equation}
that we were looking for.

Once we have dealt with the zeroth order $T|_{u=u_0}$ of the Taylor expansion of the transformation $T$, let us determine the first order (encoded in the second derivatives of the coordinate change evaluated at $u_0$; the additional terms in the expansion are irrelevant to our calculation) to implement the free-falling character for the local coordinates. For convenience in the notation, let us temporarily call $(t,r) =: (x^0,x^1)$ and $(\tl,\rl) =: (\xl^0,\xl^1)$. If $\{\xl^i,\,i=0,1\}$ is a locally free-falling coordinate system, \emph{radial} geodesic equations in these coordinates shall locally read
\begin{equation}
\derivat[2]{\xl^i}{s}{u=u_0} = 0.
\label{geod_xi}
\end{equation}
In the original Schwarzschild coordinates $\{x^i,\,i=0,1\}$ the geodesic condition becomes
\begin{equation}
\deriv[2]{x^i}{s} + \pderiv{x^i}{\xl^n} \frac{\partial^2 \xl^n}{\partial x^j \partial x^k} \deriv{x^j}{s} \deriv{x^k}{s} = 0.
\label{geod_mix}
\end{equation}
But we know that the radial geodesic equations in these coordinates must read
\begin{equation}
\deriv[2]{x^i}{s} + \Gamma^i_{j k} \deriv{x^j}{s} \deriv{x^k}{s} = 0,
\label{geod_x}
\end{equation}
where $\Gamma^i_{j k}$ are the Christoffel symbol  of the metric~\eref{metric}. Therefore, we can identify
\begin{equation}
\pderiv{x^i}{\xl^n} \frac{\partial^2 \xl^n}{\partial x^j \partial x^k} = \Gamma^i_{j k},
\label{chris_eq}
\end{equation}
and solving for the second order derivatives,
\begin{equation}
\frac{\partial^2 \xl^i}{\partial x^j \partial x^k} = \pderiv{\xl^i}{x^n} \Gamma^n_{j k}.
\label{sec_deriv_eq}
\end{equation}
First order derivatives can be read from~\eref{first_order}, and the Christoffel symbol can be computed using~\eref{metric}, finding
\begin{equation}
\eqalign{
\Gamma^r_{r r} = -\Gamma^t_{r t} = -\Gamma^t_{t r} =   - \frac{m}{r_0^2} \fr{1-\frac{2m}{r_0}},  \\
\Gamma^r_{t t} = \frac{m}{r_0^2} \left(1-\frac{2m}{r_0}\right),}
\label{christoffel}
\end{equation}
the rest of them being zero (remember that all expressions need to be evaluated at $u=u_0$). Then, second order derivatives of the coordinate transformation read
\begin{equation}
\eqalign{
\pderiv[2]{\rl}{r} =  -\frac{m}{r_0^2} \frac{\sqrt{1-\frac{2m}{r_0}+v_0^2}}{(1-\frac{2m}{r_0})^2}, \quad & \pderiv[2]{\tl}{r} =  \frac{m}{r_0^2} \frac{v_0}{(1-\frac{2m}{r_0})^2},  \\
\frac{\partial^2 \rl}{\partial r \partial t} =  -\frac{m}{r_0^2} \frac{v_0}{1-\frac{2m}{r_0}},  & \frac{\partial^2 \tl}{\partial r \partial t} =  \frac{m}{r_0^2} \frac{\sqrt{1-\frac{2m}{r_0}+v_0^2}}{1-\frac{2m}{r_0}},  \\
\pderiv[2]{\rl}{t} =  \frac{m}{r_0^2} \sqrt{1-\frac{2m}{r_0}+v_0^2},  & \pderiv[2]{\tl}{t} =  -\frac{m}{r_0^2} v_0.}
\label{sec_deriv}
\end{equation}

Having described the coordinate transformation locally up to the  first  order, we are now able to compute the acceleration of the observer~\observer as measured in the locally Minkowskian coordinate system. First, we can see using~\eref{first_order} that
\begin{equation}
\evat{\left(\deriv{\tl}{u},\deriv{\rl}{u}\right)}{u=u_0} = (1,0),
\label{com_cond}
\end{equation}
as should happen because we are in a comoving coordinate system. This means that
\begin{eqnarray}
\fl \derivat[2]{\rl}{\tl}{u=u_0} = \derivat[2]{\rl}{u}{u=u_0} = \left[ \pderiv{\rl}{r} \deriv[2]{r}{u} + \pderiv[2]{\rl}{r} \left(\deriv{r}{u}\right)^2 + \pderiv{\rl}{t} \deriv[2]{t}{u} \right. \nonumber \\
\left. + \pderiv[2]{\rl}{t} \left(\deriv{t}{u}\right)^2 + 2 \frac{\partial^2 \rl}{\partial r \partial t} \deriv{r}{u} \deriv{t}{u} \right]_{u=u_0}.
\label{acc_step}
\end{eqnarray}
The evaluation finally yields the acceleration of the trajectory of \observer in a free-falling comoving coordinate system:
\begin{equation}
\derivat[2]{\rl}{\tl}{u=u_0} = \fr{\sqrt{1-\frac{2m}{r_0}+v_0^2}} \left(a_0 + \frac{m}{r_0^2} \right) = \apr.
\label{acc}
\end{equation}
We have then justified the physical character of $\apr$.

\subsection{Physical interpretation}

As we already mentioned, expression~\eref{local_kappa} for the value of $\kp$ can be clearly interpreted physically. Let us explain each factor appearing on it. 

First of all, it contains the information corresponding to the vacuum state in $\ks$, which, as we explained, could be non-stationary. From this $\ks$ a contribution $m/r^2$ is subtracted, which is the radial acceleration due to gravitational field of the black hole.

This subtraction is multiplied by two factors. One of them is the gravitational blue-shift factor $1/\sqrt{1-2m/r}$, that appears because radiation gets red-shifted when escaping from black hole's gravity. This results in a blue-shift  for observers  near the black hole with respect to those at infinity. The other factor $\sqrt{(1-\vl)/(1+\vl)}$ is clearly a Doppler shift factor. The velocity that appears in this factor is $\vl$, which is the velocity of the observer~\observer with respect to the black hole.  It is measured in a locally Minkowskian coordinate system [see~\eref{first_order}], in which the speed of light is always equal to $1$.

Finally, the last term $\apr$ is just the proper acceleration of the observer~\observer. Being measured with respect to an instantaneously comoving and free-falling observer, it is the acceleration that the observer~\observer would actually feel whenever switching on its rockets. When $\kp$ can be interpreted as a strict temperature, this term corresponds precisely to Unruh thermal radiation (temperature proportional to the proper acceleration), which is present even in Minkowski spacetime for accelerated observers~\cite{Unruh:1976db}. It is remarkable that this proper acceleration appears in the expression directly as an additive quantity. 

We remind that expression~\eref{local_kappa} is \emph{always exact,} no matter what interpretation one can assign to $\kp$ itself. When  condition~\eref{adiabatic_condition_alt} is satisfied, and one can talk about a strict thermal perception with temperature proportional to $\kp$, the interpretation of expression~\eref{local_kappa} is literal in terms of physical phenomena dealing with Hawking radiation which is modified by gravitational blue-shifts, Doppler shifts and Unruh radiation. Otherwise, although exact, it is only a geometrical estimator of the amount of particle perception. A more detailed analysis of particle perception would require the use of models of detectors and calculation of response functions~\cite{Louko:2007mu}.

\section{Some applications of the formula\label{applications}}

To end this work, let us put in practice formula~\eref{local_kappa} with some examples of observers following different trajectories outside the black hole. In~\sref{kappa_state}, we used expression~\eref{ks_def} to study the properties of different vacuum states regardless of the observer. In the same way, we can now use formula~\eref{local_kappa} to study the perception of different observers without the need of choosing a particular vacuum state. The particulars of this choice will remain encoded in the generic function $\ks$.

For static observers at a fixed radius $r(u) = \rc = \cte$, we clearly have 
\begin{equation}
\vl(u) = 0, \quad \apr(u) = \fr{\sqrt{1-\frac{2m}{\rc}}}\frac{m}{\rc^2},
\label{static_prop}
\end{equation}
and thus
\begin{equation}
\kp (u) = \fr{\sqrt{1-\frac{2m}{\rc}}} \ks(\ub(u)).
\label{static_kp}
\end{equation}
These observers are perceiving the radiation escaping from the black hole at this moment, multiplied by the gravitational blue-shift factor associated with their radial position. The fulfilment of the adiabatic condition~\eref{adiabatic_condition_alt} will depend on the particular vacuum state. For stationary vacuum states, the effective temperature will be constant, and the adiabatic condition will be perfectly satisfied. For non-stationary states, it will depend on its particular evolution.

In~\cite{Barbado:2011dx} we defined an \emph{Unruh observer} as one which is instantaneously static, let us say at $u=u_0$, and free-falling. In this case, we have $\vl (u_0) = 0$ and $\apr (u_0) = 0$, and thus
\begin{equation}
\kp (u_0) = \fr{\sqrt{1-\frac{2m}{r(u_0)}}} \left( \ks(\ub(u_0)) - \frac{m}{r(u_0)^2} \right).
\label{unruh_kp}
\end{equation}
In particular, consider the Unruh vacuum state for the field so that $\ks = 1/(4m)$. When the Unruh observer as defined above is arbitrarily near the event horizon, we have
\begin{equation}
\fr{\sqrt{1-\frac{2m}{r(u_0)}}} \left( \fr{4m} - \frac{m}{r(u_0)^2} \right) \sra[r(u_0) \to 2m] 0.
\label{unruh_limit}
\end{equation}
This agrees with the result found by Unruh that free-falling observers at the horizon perceive no radiation~\cite{Unruh:1976db}. However, as was discussed thoroughly in~\cite{Barbado:2011dx}, this is only true for observers which are also instantaneously static.

Let us consider the perception of a general free-falling observer when crossing the event horizon. For simplicity, we will assume again the Unruh vacuum state. The now non-vanishing radial velocity pointing towards the black hole, $v := \hderiv{r}{u} < 0$, implies that the velocity $v_l$ behaves as
\begin{equation}
\vl = \frac{v}{\sqrt{1-\frac{2m}{r}+v^2}} \sra[r \to 2m] -1.
\label{free-falling_vs}
\end{equation}
But this implies a diverging Doppler shift
\begin{equation}
\sqrt{\frac{1-\vl}{1+\vl}} \sra[r \to 2m] \infty,
\label{free-falling_Ds_infty}
\end{equation}
that will compete with the vanishing of~\eref{unruh_limit}. This competition leads to a finite result for $\kp$ at the event horizon. In particular, if the observer trajectory is that of a free-falling observer coming from radial infinity, it can be shown~\cite{Barbado:2011dx}  that
\begin{equation}
\kp \sra[r \to 2m] \fr{m}.
\label{free-falling_kp_infty}
\end{equation}
That is, the observer will perceive an effective-temperature function four times bigger than the value of $1/(4m)$ of the usual Hawking radiation.  As for the adiabatic condition~\eref{adiabatic_condition_alt}, one can also show \cite{Barbado:2011dx} that $\epsilon \to 3/8$ when $r \to 2m$. Thus, the adiabatic condition is not strictly satisfied, but the value of $\epsilon$ is still smaller than $1$. We can conclude that, although not perfectly thermal, the spectrum perceived when crossing the horizon will be not very different from thermal. In any case, the observers will perceive a non-zero flux of particles.

At present the authors are using the potentiality of this formula to analyse buoyancy effects in black holes.

\ack

The authors want to thank an anonymous referee for some useful comments and suggestions. Financial support was provided by the Spanish MICINN through the projects FIS2008-06078-C03-01, FIS2008-06078-C03-03, FIS2011-30145-C03-01 and FIS2011-30145-C03-02 (with FEDER contribution), and by the Junta de Andaluc\'{\i}a through the project FQM219.

\section*{References}

\bibliography{barb2012}
\bibliographystyle{unsrt}

\end{document}